\documentclass{midl}

\usepackage{booktabs}
\usepackage{graphics}
\usepackage{multirow}
\usepackage{todonotes}
\usepackage[skip=2.5pt]{caption}

\jmlrproceedings{MIDL}{Medical Imaging with Deep Learning}
\jmlrpages{}
\jmlryear{2019}
\jmlrworkshop{MIDL 2019 -- Extended Abstract Track}

\title[Conditioning Convolutional Segmentation Architectures with Non-Imaging Data]{Conditioning Convolutional Segmentation Architectures with Non-Imaging Data}

\midlauthor{\Name{Grzegorz Jacenk\'ow\nametag{$^{1}$}} \Email{g.jacenkow@ed.ac.uk} \\
            \Name{Agisilaos Chartsias\nametag{$^{1}$}} \\
            \Name{Brian Mohr\nametag{$^{2}$}} \\
            \Name{Sotirios A. Tsaftaris\nametag{$^{1, 3}$}} \\
            \addr $^{1}$ The University of Edinburgh, Edinburgh, United Kingdom \\
            \addr $^{2}$ Canon Medical Research Europe Ltd., Edinburgh, United Kingdom \\
            \addr $^{3}$ The Alan Turing Institute, London, United Kingdom
}

\begin{document}

\maketitle

\begin{abstract}
We compare two conditioning mechanisms based on concatenation and feature-wise
modulation to integrate non-imaging information into convolutional neural
networks for segmentation of anatomical structures. As a proof-of-concept we
provide the distribution of class labels obtained from ground truth masks to
ensure strong correlation between the conditioning data and the segmentation
maps. We evaluate the methods on the ACDC dataset, and show that conditioning
with non-imaging data improves performance of the segmentation networks. We
observed conditioning the U-Net architectures was challenging, where no method
gave significant improvement. However, the same architecture without skip
connections outperforms the baseline with feature-wise modulation, and the
relative performance increases as the training size decreases.
\end{abstract}


\section{Introduction}
Integrating non-imaging modalities becomes of interest to the research community
with dedicated workshops such as beyondMIC 2018. The majority of the work has
been focused on multi-modal data fusion where each modality is mapped to an
embedding otherwise by concatenating~\cite{Tiwari2011} or maximising correlation
between the views~\cite{Golugula2011}. Other approaches include intermediate
functions such as neural networks combined with a linear classifier. For
instance, \cite{Shmulev2018} developed a method to predict conversion of mild
cognitive impairment to Alzheimer's disease, and \cite{Cerna2019} shown a
classifier to estimate probability of mortality within one year. However, to the
best of our knowledge the current techniques do not apply to the segmentation
networks as they focus on classification.

In this work, we use segmentation of Cardiovascular Magnetic Resonance (CMR)
images as an example for evaluating the conditioning mechanisms. We observe that
the majority of the proposed approaches for CMR segmentation rely only on imaging data, and do not
incorporate additional information available in Electronic Health Records
(EHRs)~\cite{Bizopoulos2019}. As collecting such information is often more
time-efficient than the annotation process, we motivate to investigate how
non-imaging data can be used as prior in convolutional segmentation networks. To
ensure strong correlation of the conditioning information with the segmentation
task, and to avoid inter-subject bias due to pathologies, we propose a
proof-of-concept where the network is conditioned on the distribution of class
labels obtained from the ground truth masks. Together with the image, we provide
to the networks expected percentage of pixels in the output mask corresponding
to each class, i.e. myocardiun, left- and right ventricular cavities. This
conditioning data is an approximation of the heart's size, which is a common
biomarker easily extracted from echocardiography images~\cite{Jenkins2008}.

\section{Methodology}
We consider concatenation-based conditioning and feature-wise modulation applied
to two networks for 2D segmentation; a U-Net~\cite{Ronneberger2015}
where each convolutional layer is followed by batch
normalisation~\cite{Mousavi2016}, and an encoder-decoder that has the same
architecture as the U-Net except there are no skip connections.

\noindent
\textbf{Concatenation-based conditioning} refers to methods where the
conditioning information is concatenated with a feature map or with the model's
input. We evaluate two approaches for acquiring the conditioning representation
$\tilde{z}$. Given a distribution of class labels $z$~\footnote{To address the
class imbalance, we exclude background labels and multiple the other classes by
100.} and a function $f$, $\tilde{z} = f(z)$, where $f$ is either an identity
function (referred as \textit{raw concatenation}) or a fully-connected
3-6-12-6-3 network (referred as \textit{MLP concatenation}). We apply the
concatenation-based conditioning at three levels: \textit{early fusion} with
spatial replication of the input-level features, \textit{middle fusion} at the
latent space of the encoder-decoder networks, and \textit{late fusion} before
the last convolutional layer.

\noindent
\textbf{Feature-wise Linear Modulation} (FiLM)~\cite{Perez} can be classified as
an instance normalisation~\cite{Ulyanov2016} technique in which a scaling
$\gamma$ and a shifting $\beta$ factors are applied to a particular channel $c$
in a feature map $F_c$, i.e. $\textrm{FiLM}(F_c | \gamma_c, \beta_c) = \gamma_c
F_c + \beta_c$. In contrast to the regular instance normalisation, the factors
are learnt with a multilayer perceptron from an input $z$. Our work focuses on
applying FiLM layers along the decoder path (\textit{decoder fusion}) and before
the final convolutional layer (\textit{late fusion}).

\section{Experiments}
\textbf{Dataset.} We use the cardiac cine-MRI dataset from the ACDC 2017
challenge~\cite{Bernard2018} for the task of segmenting the images into three
anatomical structures, i.e. myocardium, left- and right ventricular cavities.
The annotated dataset contains images at end-systolic and -diastolic phases from
100 patients, and varying spatial resolutions. We resample the volumes to a
common resolution of 1.37 $\textrm{mm}^2$ per pixel, resize each slice to 224 x
244 pixels, and standardise intensities using z-score with clipping values
exceeding three units of standard deviation from the volume's mean.

\noindent
\textbf{Training and Evaluation.} All models are trained using
Adam~\cite{Kingma2014} optimiser with learning rate $\alpha = 0.0001$, and Focal
Loss~\cite{Lin2017} with $\gamma = 0.5$. The networks are trained with 500
epochs and we apply early stopping with patience set to 100. To determine the
effect of conditioning mechanisms on datasets with limited amount of training
examples, we repeat each experiment with varying fractions of the training set,
i.e. at 100\%, 25\%, 6\% and 1.5\% (single subject). The experiments are
evaluated using 3-fold cross validation with subjects shuffled and split into
70\%, 15\%, 15\% training, validation and test sets respectively. We calculate a
Dice score for 3D volumes as the average across all anatomical
structures. We report average Dice, standard deviation and evaluate statistical
significance (5\%) using paired t-test with Bonferroni correction on the test
sets as we compare each conditioning mechanism with the corresponding baseline,
i.e. we make 8 comparisons.

\noindent
\textbf{Results.} The empirical results of the U-Net and the encoder-decoder
networks are presented in Table~\ref{table:unet-results} and
Table~\ref{table:encoder_decoder-results} respectively. Overall, it can be seen
that conditioning on non-imaging information improves segmentation performance
in terms of Dice. In particular future-wise modulation has relative improvement
of 2\% - 19\% over the encoder-decoder baseline. We also observe that relative
performance increases as the size of the training dataset decreases. However,
the U-Net architectures show to be more challenging for integrating non-imaging
information. Although, concatenation of raw values before the last convolutional
layer has outperformed the U-Net baseline by a margin of 1\% - 51\%, the
variance remains high. Furthermore, the results do not show significant relative
improvement with limited training examples as in the encoder-decoder networks.

\begin{table}
\centering
    \resizebox{\columnwidth}{!}{%
        \begin{tabular}{|c||c|c|c|c|c|c|c|c|c|}
        \hline
        \multirow{2}{*}{Fraction} & \multirow{2}{*}{Baseline} & \multicolumn{3}{c|}{Concatenation (raw)} & \multicolumn{3}{c|}{Concatenation (MLP)} & \multicolumn{2}{c|}{FiLM} \\
                                  &                           & Early & Middle & Late                   & Early & Middle & Late                   & Decoder & Late \\
        \hline
        100\% & $.89_{\pm .04}$ & $.89_{\pm .03}$ & $.90^*_{\pm .03}$ & $.90_{\pm .04}$ & $.89_{\pm .03}$ & $.90_{\pm .03}$ & $.90_{\pm .03}$ & $\textbf{.91}^*_{\pm .02}$ & $.90^*_{\pm .03}$ \\
        \hline
        25\% & $.80_{\pm .13}$ & $.81_{\pm .10}$ & $.82_{\pm .09}$ & $\textbf{.83}^*_{\pm .10}$ & $.81_{\pm .11}$ & $.82_{\pm .10}$ & $.82_{\pm .10}$ & $.81_{\pm .12}$ & $.82_{\pm .10}$ \\
        \hline
        6\% & $.39_{\pm .29}$ & $.53^*_{\pm .23}$ & $.58^*_{\pm .27}$ & $\textbf{.59}^*_{\pm .25}$ & $.58^*_{\pm .28}$ & $.54^*_{\pm .26}$ & $.58^*_{\pm .25}$ & $.55^*_{\pm .26}$ & $.55^*_{\pm .26}$ \\
        \hline
        1.5\% & $.41_{\pm .23}$ & $.42_{\pm .23}$ & $\textbf{.44}_{\pm .22}$ & $.42_{\pm .22}$ & $\textbf{.44}_{\pm .24}$ & $.42_{\pm .22}$ & $.42_{\pm .22}$ & $.38_{\pm .24}$ & $.34^*_{\pm .23}$ \\
        \hline
        \end{tabular}%
    }
\caption{Performance of the networks with U-Net architecture as an average over
         Dice scores for LVC, myocardium and RVC. The best results are shown in
         \textbf{bold}. An asterisk (*) denotes the statistical significance
         (5\%) comparing to the baseline.}
\label{table:unet-results}
\end{table}

\begin{table}
    \resizebox{\columnwidth}{!}{%
        \begin{tabular}{|c||c|c|c|c|c|c|c|c|c|}
        \hline
        \multirow{2}{*}{Fraction} & \multirow{2}{*}{Baseline} & \multicolumn{3}{c|}{Concatenation (raw)} & \multicolumn{3}{c|}{Concatenation (MLP)} & \multicolumn{2}{c|}{FiLM} \\
                                  &                           & Early & Middle & Late                   & Early & Middle & Late                   & Decoder & Late \\
        \hline
        100\% & $.87_{\pm .04}$ & $.86^*_{\pm .04}$ & $.88^*_{\pm .03}$ & $.88^*_{\pm .03}$ & $.87_{\pm .04}$ & $.88^*_{\pm .03}$ & $.87^*_{\pm .03}$ & $\textbf{.89}^*_{\pm .02}$ & $.88^*_{\pm .03}$ \\
        \hline
        25\% & $.78_{\pm .09}$ & $.75^*_{\pm .11}$ & $.75^*_{\pm .12}$ & $.78_{\pm .09}$ & $.78_{\pm .10}$ & $.76_{\pm .11}$ & $.77_{\pm .10}$ & $\textbf{.82}^*_{\pm .06}$ & $.78_{\pm .10}$ \\
        \hline
        6\% & $.55_{\pm .22}$ & $.53_{\pm .22}$ & $.53_{\pm .22}$ & $.55_{\pm .23}$ & $.52^*_{\pm .23}$ & $.51^*_{\pm .23}$ & $.54_{\pm .22}$ & $\textbf{.58}_{\pm .19}$ & $.48^*_{\pm .24}$ \\
        \hline
        1.5\% & $.31_{\pm .17}$ & $.34_{\pm .19}$ & $.31_{\pm .17}$ & $.31_{\pm .19}$ & $\textbf{.38}^*_{\pm .21}$ & $.29_{\pm .18}$ & $.35^*_{\pm .19}$ & $.37^*_{\pm .18}$ & $.31_{\pm .18}$ \\
        \hline
        \end{tabular}
    }
\caption{Same as Table~\ref{table:unet-results} but for the encoder-decoder architecture.}
\label{table:encoder_decoder-results}
\end{table}

\section{Conclusion}
We have considered the task of conditioning segmentation networks on non-imaging
data. We have shown that conditioning with non-imaging data improves
performance of the segmentation networks with feature-wise modulation for the
encoder-decoder networks yielding a consistent improvement. However, conditioning
the U-Net networks is challenging and the same methods do not result in significant improvement.

\midlacknowledgments{This project was supported by the Royal Academy of
Engineering under the Research Chairs and Senior Research Fellowships scheme. We
also thank the School of Engineering at The University of Edinburgh and Canon
Medical Research Europe Ltd. for the scholarship and funding.}

\end{document}